# On the dark matter's halo theoretical description

L M Chechin

*V G Fessenkov Astrophysical Institute, National Centre for cosmic researches and technologies, National Space Agency, 050020, Almaty, Kazakhstan*
*E-mails: chechin-lm@mail.ru; leonid.chechin@gmail.com*

**Abstract.** We argued that the standard field scalar potential couldn't be widely used for getting the adequate galaxies' curve lines and determining the profiles of dark matter their halo. For discovering the global properties of scalar fields that can describe the observable characteristics of dark matter on the cosmological space and time scales, we propose the simplest form of central symmetric potential celestial – mechanical type, i.e. $U(\varphi) = -\mu/\varphi$. It was shown that this potential allows get rather satisfactorily dark matter profiles and rotational curves lines for dwarf galaxies. The good agreement with some previous results, based on the N-body simulation method, was pointed out. A new possibility of dwarf galaxies' masses estimation was given, also.

**1. Introduction.** The outstanding peculiarity of modern cosmology just consists in what, that it allows carrying out the high-precision measurements of the Universe physical parameters that were considered as impossible not so long ago. Talk, for example, about the measurement of cosmic microwave background anisotropy [1], about the polarization of cosmic microwave background [2], about the gravitational microlensing [3] and some other observable effects of modern cosmology.

This peculiarity of modern cosmology allows using broadly the observable data for next development of theoretical models those processes that are the paramount for understanding the structure and evolution of the Universe. In doing this we concentrate our attention on problem that is orientating on the dark matter phenomena understanding, first of all. Some new results in this sphere have been presented in the recent article [4].

This article organized as follows. In second section we briefly describe the shape of a galaxy's halo of dark matter. In third section we consider the scalar field of the oscillator type and demonstrate that standard scalar potential couldn't be productively used for getting the adequate galaxies' curve lines and determining the profiles of dark matter halo. For doing this it is necessary choose another type of scalar potential. This problem was searched in section 4. There was argued that global properties of scalar fields is possible describe by the central symmetric potential celestial – mechanical type, i.e. by $U(\varphi) = -\dfrac{\mu}{\varphi}$. Section 5 was devoted to calculating profiles of dark matter halo and rotational curves of a galaxy on the basis of potential celestial – mechanical type. Some astronomical predictions concerning the physical characteristics of dwarf galaxies were argued in last - sixth - section.

**2. Scalar field for describing halo of the dark matter.** Talking about the dark matter it's necessary mention the pioneer Zwicky article [5], where the idea about presence an unknown type of substance (dark matter) in galaxies was put forward at first. This substance ensures the stability of any galaxy. That is why the modern vision about the galaxy structure includes the dark matter halo as its indispensable component [6].

According this as the simplest model of a galaxy the following system of number subsystems is considering – massive nuclei (the rotating black hole, usually), bulge, spherical stars shell, gaseous flat disc and halo of dark matter. Note that stars shell and flat disc have the common sizes in order of $10 Kpc$, while the typical sizes of dark matter halo are about of $100 Kpc$ and larger.

The observable data show that halo of dark matter contains the main part of galaxy's mass (about 90%). The typical total galaxy mass estimates as $M_{Gal} \sim (10^{12}\text{-}10^{14})M_{Sun}$, where $M_{Sun}$ - mass of the Sun [7].

Because the goal of our article is searching some properties of the dark matter, only, we'll neglect all of galaxy's components except the halo of dark matter in following.

In accordance with number of articles, enumerated, for example in [6], the dark matter possible describes by setting the suitable scalar field. For classical scalar field $\varphi$ the energy-momentum tensor has the form

$$T_{\mu\nu} = \partial_\mu \varphi \cdot \partial_\nu \varphi - \frac{1}{2} g_{\mu\nu} \left[ (\partial_\alpha \varphi)^2 - U(\varphi) \right], \qquad (1)$$

where $U(\varphi)$ - the potential energy of scalar field, $g_{\mu\nu}$ - the metric tensor of an external gravitational field. Later on we'll consider flat space-time, i.e. $g_{\mu\nu} = \delta_{\mu\nu}$.

Basing on (1) it is easy get the expression for the scalar field energy density

$$T_{00} = \rho = \frac{1}{2} \left[ (\partial_m \varphi)^2 + U(\varphi) \right], \qquad (2)$$

the space momentum tensor

$$T_{kl} = \partial_k \varphi \cdot \partial_l \varphi + \frac{1}{2} \left[ (\partial_m \varphi)^2 - U(\varphi) \right] \cdot \delta_{kl} \qquad (3)$$

and the expression for isotropic pressure $p = \frac{1}{2} \cdot \left[ (\partial_m \varphi)^2 - U(\varphi) \right]$.

Now we'll find the field equation by usage the law of energy-momentum conservation

$$\nabla_\nu T^{\mu\nu} = 0. \qquad (4)$$

Remembering that $g_{\mu\nu} = \delta_{\mu\nu}$ and considering the Newtonian limit, from (4) we get

$$\Delta \varphi + \frac{\partial U(\varphi)}{\partial \varphi} = 0. \qquad (5)$$

**3. Scalar field of the oscillator type.** Choosing the effective potential energy as $U(\varphi) = \frac{1}{2}m^2\varphi^2$ that corresponds to the standard scalar potential of the oscillator type, we get the well-known equation

$$\Delta\varphi + m^2\varphi = 0. \tag{6}$$

Passing to the new variable $\psi = \varphi \cdot r$ and using the spherical coordinates, (6) rewrites as

$$\frac{d^2\psi}{d\tilde{r}^2} + \psi = 0, \tag{7}$$

where $\tilde{r} = mr$ is the dimensionless variable. It has the standard solution of oscillator type

$$\psi = \psi_0 \exp(i\tilde{r}) = \psi_0 \exp(imr). \tag{8}$$

Now the question arises – what of magnitude the variable $\tilde{r} = mr$ may be? For its estimating remember that minimal mass $m$ of the scalar field particles, describes the cold dark matter, according [8], equals $10^{-23} eV$. Then in usual units ($1 eV \approx 1,8 \cdot 10(-33) g$; $\hbar \approx 1,0 \cdot 10(-27) erg \cdot \sec$; $c \approx 3,.0 \cdot 10(10) cm \cdot \sec^{-1}$) the maximal corresponding distance will be

$$r_{max} \approx \frac{\hbar}{c} m^{-1} \sim 10 pc = 0,1 Kpc. \tag{9}$$

But from the set of observable data the most interesting distances for a galaxy's curve lines is about ten and more kiloparsecs. Hence, $\tilde{r}$ is the large parameter satisfies following condition $\tilde{r} = mr = \frac{r}{r_{max}} \gg 1$. Substitution (8) and its derivative into (2) give the explicit and exact form of dark matter density profile

$$\rho(r) = \frac{\psi_0^2}{2r^2} \left[ \left( \frac{1}{r^2} - \frac{1}{r_{max}^2} \right) + \frac{1}{r_{max}^2} \right] \exp(2imr). \tag{10}$$

From expression (10) anyone easy gets the scalar field mass density

$$\rho(r) = \frac{m^2 \psi_0^2}{2} \cdot \frac{\cos^2 mr}{m^2 r^4} = \rho_0 \cdot \frac{\cos^2 mr}{m^2 r^4}. \tag{11}$$

From all of above we may write down the Poisson equation for a searching gravitational potential $\Pi(r)$ of the scalar field $\varphi$ in spherical coordinates

$$\frac{1}{r^2} \cdot \frac{d}{dr}\left( r^2 \frac{d\Pi(r)}{dr} \right) = -4\pi G \rho_0 \cdot \frac{\cos^2 mr}{m^2 r^2}. \tag{12}$$

From this equation after its first integration and omitting the periodical terms it follows the functional dependency

$$\frac{d\Pi(r)}{dr} \sim -\frac{1}{r^3} \qquad (13)$$

and after second integration we find the potential space dependency

$$\Pi(r) \sim -\frac{1}{r^2} \qquad (14)$$

Expression (13) allows find the rotational curve line by equating it to the specific centrifugal force, i.e. $\frac{v^2}{r} \sim \frac{1}{r^3}$. So, the rotational curve line has the next functional dependency

$$v(r) \sim \frac{1}{r}. \qquad (15)$$

Now question arises - how does the curve line (15) correlate with the real observable data? Basing on the observable results of dark and baryonic matter distributions in 34 bright spiral galaxies [9], we may conclude the follows.

First, there are number of variants that modeling the density distribution of dark matter halos by N-body simulation method [10]. The most part of them predicts satisfactory that dark matter is essential to the inner radii of galaxies, i.e. for the regions where $\frac{r}{r_{max}} \sim 1$, in contradiction to maximal sizes of halo which we are considering previously.

Second, the observable rotational curves don't represent by the curve lines type of hyperbola. Hence, expression (15) is the result of poorly theoretical determined dark matter distribution even in galaxies where the presence of dark matter is dominant.

From all of sad above it is clear, that standard scalar potential couldn't be productively used for getting the adequate galaxies' curve lines and determining the profiles of dark matter halo. For doing this it is necessary to choose another type of scalar potential. Note that Yukawa-type potential for describing dark matter has been considered by A. Loeb and N. Weiner [11] recently.

**4. Scalar field of the celestial – mechanical type.** In fact, the potentials type of $U(\varphi) = \frac{1}{2}m^2\varphi^2$ and analogous them (self-acting potentials, Higgs potentials, Yukawa-type potential, etc) are using for searching the local properties of the scalar fields on small space and time intervals (for very early and early Universe). But our aim is discovering the global properties of scalar fields that can describe the observable characteristics of dark matter on the cosmological space and time scales.

The global properties of scalar fields, as it seems, is possible describe by the simplest form of central symmetric potential celestial – mechanical type, i.e. by $U(\varphi) = -\frac{\mu}{\varphi}$. Thus the corresponding unit Lagrangian takes on the form

$$L = \frac{1}{2}(\vec{\nabla}\varphi)^2 + \frac{\mu}{\varphi}, \qquad (16)$$

where $\mu$ is any constant value. Its possible interpretation sees in section 6. For deducing (16) we based on the Lagrangian of a probe particle moving in the central symmetric field and used the standard formal replacing $t \to x^k$ and $r = \sqrt{(x^k)^2} \to \varphi$, that usually applies in the field theory [12].

Earlier was pointed out that simplest halo of dark matter have the spherical – symmetric shape, i.e. it depends from radius $r$ only. That is why consider the one-dimensional operator nabla $\vec{\nabla} \to \frac{d}{dr}(\ )\cdot\vec{e}_r$. So, in place of (6) we get the following equation in the dimensionless values $r$ and $\varphi$ (as in (7))

$$\frac{d^2\varphi}{dr^2} + \frac{\mu}{\varphi^2} = 0, \qquad (17)$$

that describes the linear "movement" of a unit mass particle's in the central field. The angular momentum $M$ for such type of "movement", as it well known, equals to zero, i.e. we may set $M = 0$. Hence, for finding $\varphi(r)$ and its first derivative $\frac{d\varphi(r)}{dr}$ it is preferring use Lagrangian (16) than equation of "motion" (17). According classical textbooks [13] we have

$$r = \int \frac{d\varphi}{\sqrt{2\left(E + \frac{\mu}{\varphi}\right)}}. \qquad (18)$$

Now consider the field analogs of some types of movement that are interesting from dynamical viewpoint and having width applications in the celestial mechanics.

i) Assume that "total energy" is more larger than "potential energy", i.e. that $E \gg \frac{\mu}{\varphi}$. From physical viewpoint this condition means choosing the space region of dark matter halo that closed to center of a galaxy. Then integral (18) takes on the approximate expression (integrating constant have been included into $r$)

$$r = \frac{1}{\sqrt{2E}}\left(\varphi - \frac{\mu}{2E}\ln\varphi\right). \qquad (19)$$

For its inversing let $r = r_0 + \delta r$ and $\varphi = \varphi_0 + \delta\varphi$, where $r_0$ and $\varphi_0$ are the main terms, while $\delta r$ and $\delta\varphi$ are the small additives order of $\frac{\mu}{E\varphi_0} \ll 1$ to them. That is why $\varphi_0 = \sqrt{2E \cdot r_0}$ and $\delta\varphi = \sqrt{2E} \cdot \delta r + \frac{\mu}{2E}\ln\varphi_0$. These expressions allow write down the final result in the following form and with the mentioned above accuracy

$$\varphi(r) = \sqrt{2E} \cdot r + \frac{\mu}{2E}\ln\left(\sqrt{2E} \cdot r\right). \qquad (20)$$

ii) Let the opposite correlation $E \ll \frac{\mu}{\varphi}$ takes place. This condition, contrary to previous one, means choosing the space region far from center of galaxy. Then from (18) we get immediately (integrating constant have been included into $r$, also)

$$r = \frac{1}{\sqrt{\mu}}\left(\frac{2}{3}\varphi^{3/2} + \frac{1}{2}\frac{E}{\mu}\varphi^{5/2}\right). \tag{21}$$

In full analogy with the first variant we set $r = r_0 + \delta r$ and $\varphi = \varphi_0 + \delta\varphi$, where $r_0$ and $\varphi_0$ are the main terms, whereas $\delta r$ and $\delta\varphi$ are the small additives order of $\frac{\mu}{E\varphi_0} \gg 1$ to them. Such representation leads to the next approximate functional dependency

$$\varphi(r) = (\frac{3}{2}\sqrt{\mu})^{2/3} \cdot r^{2/3} + \frac{1}{2} \cdot (\frac{3}{2})^2 \cdot \frac{E}{\sqrt[3]{\mu}} \cdot r^{4/3}. \tag{22}$$

So, expressions (20) and (22) describe explicitly the scalar field potential in the whole region – closed to the center and far from the center – of dark matter halo distribution.

**5. Profiles of dark matter halo and rotational curves of a galaxy.** Profile of dark matter is important characteristic of a galaxy structure. It allows calculate the corresponding gravitational potential, find the galaxy's rotational curve and make some other cosmological conclusions. In most articles the profiles of dark matter halo and the rotational curves of a galaxy were searched by the N-body simulation method, mainly. But we'll consider them from the theoretical field viewpoint.

Now, basing on (20) and (22), it possible finds profiles of dark matter halo and rotational curve lines for any galaxy.

i) First of all it is easy calculate the first derivatives from potential (20) -

$$\frac{d\varphi(r)}{dr} = \sqrt{2E} + \frac{\mu}{\sqrt{2E}} \cdot \frac{1}{r}. \tag{23}$$

Substituting (20) and (23) into (2) we get the following mass density profile of dark matter

$$\rho(r) = \frac{1}{2}\left[\left(1 + \frac{1}{\sqrt{2E}}\right)\frac{\mu}{r} - \left(\frac{\mu}{2E}\right)^2 \frac{1}{r}\ln\sqrt{2E} \cdot r\right]. \tag{24}$$

The integrating constant has been included into the left side of (24). Examining it we see that mass density of dark matter decreases in space and becomes equal to zero at the distance

$$\sqrt{2E} \cdot r_0 = \exp\left((2E)^2\left(1 + \frac{1}{\sqrt{2E}}\right)/\mu\right). \tag{25}$$

At distances $r > r_0$ the mass density of dark matter becomes growth again. This circumstance is very important, because in the neighboring region the rotational curve line must possess by small "gap" that is possible observing in principle.

Results of articles [14] gives that authors' mean radial density profile at small distances decreases as $\rho(r) \sim \frac{1}{r}$ that is in good correlation with our estimation (24).

Now remembering that ratios $\frac{\mu}{E\varphi_0} \ll 1$ and $\frac{\mu}{E^{3/2}}\frac{1}{r} \ll 1$ take places, it possible omits the second term in (24) as the value of larger order of minuteness. Our next step is finding the gravitational potential of field produced by main term in mass density (24). The Poisson equation in the spherical coordinates takes on the form

$$\frac{1}{r^2}\frac{d}{dr}\left(r^2 \frac{d\Pi(r)}{dr}\right) = -2\pi G\left(1 + \frac{1}{\sqrt{2E}}\right)\frac{\mu}{r} = -2\pi G \frac{\tilde{\mu}}{r}. \qquad (26)$$

Integration (26) gives

$$\frac{d\Pi(r)}{dr} = -2\pi G \tilde{\mu} = C = const \qquad (27)$$

and

$$\Pi(r) = C \cdot r, \qquad (28)$$

hence.

From (27) easy get the rotational curve line by its equating to the centrifugal force that acts at a "probe" star. Thus, the following relation

$$\frac{v^2}{r} = C \qquad (29)$$

and the corresponding simplest rotational curve line

$$v(r) \sim r^{1/2}, \qquad (30)$$

that analogously (15) is the direct consequence of (29), take places. From the geometrical viewpoint in coordinates $\{v;r\}$ this curve represents the line that slowly growth with distance increasing. That is why the shape of line (30) is in good correlation with the real rotational curve lines, because starting from any distance all of them possess by a weak trend for growth with the distance increasing. The results of dark matter's profiles observing at short distances (~ $30 Kpc$), that are proofing our conclusion, present in article [9].

ii) In full analogy with the previous variant we calculate the first derivative of (22)

$$\frac{d\varphi(r)}{dr} = \frac{2}{3}\left(\frac{3}{2}\sqrt{\mu}\right)^{2/3} \cdot r^{-1/3} + \frac{2}{3}\left(\frac{3}{2}\right)^2 \frac{E}{\sqrt[3]{\mu}} \cdot r^{1/3}. \qquad (31)$$

Substituting (22) and (31) into (2) and remembering that $\frac{E\varphi_0}{\mu} \ll 1$, analogously to our foregoing result, the approximated expression takes place

$$\rho(r) = \rho_0 + \left[ \frac{2}{9} \left( \frac{3}{2} \sqrt{\mu} \right)^{4/3} + \frac{2\tilde{\mu}}{\left(\frac{3}{2}\right)^2} \frac{1}{\sqrt[3]{\mu}} \right] \cdot r^{-2/3} = \rho_0 + \sigma_0 \cdot r^{-2/3}, \qquad (32)$$

where $\rho_0$ is the corresponding small (order of $\frac{E\varphi_0}{\mu} \ll 1$) constant value.

Examining article [14], cited above, we see that author's mean radial density profile at large distances, contrary to our result (32), decreases more rapidly, namely as $\rho(r) \sim \frac{1}{r^{5/2}}$. The same conclusion follows from article [9], where the mass density profile changes according the dependency $\rho(r) \sim \frac{1}{r^{31/9}}$

But at the same time dark matter profile (32) is in good correlation with the results of article [15], where the shape of profile density for dark matter dominated dwarf and low-surface brightness late-type galaxies describes as $\rho(r) \sim \frac{1}{r^\gamma}$, $\gamma \approx 0.2 - 0.4$. It is important to underline that instead of other articles the power index in profile density for such type of cosmic objects is smaller than unite.

Now, omitting in (32) the constant term we get equality for finding the gravitational potential. In fact, the Poisson equation takes on the form

$$\frac{1}{r^2} \frac{d}{dr}\left( r^2 \frac{d\Pi(r)}{dr} \right) = -4\pi G \frac{\sigma_0}{r^{2/3}}. \qquad (33)$$

Then

$$\frac{d\Pi(r)}{dr} = -4\pi G \sigma_0 \cdot r^{1/3}. \qquad (34)$$

From (34) we get the rotational curve line by its equating to the centrifugal force. Thus we have the needed relation

$$\frac{v^2}{r} = 4\pi G \sigma_0 \cdot r^{1/3} \qquad (35)$$

and the corresponding curve line

$$v(r) \sim r^{2/3}. \qquad (36)$$

From the geometrical viewpoint in coordinates $\{v; r\}$ curve line (36) represents the line that growth with distance increasing more rapidly than curve (30). That is why its shape is differing from the real rotational curve lines. But it is necessary remember that theoretical curve line relates to large distances ($\frac{\mu}{\varphi} \gg E$), while all of having now observable curve lines were plotted for the short distances (about $30-40 Kpc$). And now we once again are quoting the above mentioned article by Kravtsov and Klypin [16], where the rotational curve reproduces by dependency $v(r) \sim r^q$ with $q \approx 0.9 \div 0.8$. This power index coincides practically with our result (36).

Note that behavior of curve lines (30) and (36), if they apply to galaxies movement, is also in good correlation with the results of searching hidden mass in the Local Group [17].

**6. Conclusion.** From body of article we may conclude the following. Our choice of scalar field the celestial – mechanical type allowed productively describes the dark matter halo's profiles at small and large distances. They are in rather good correlation with results of Kravtsov and Klypin on the modeling of profile density and rotational curve lines shapes for the dark matter dominated dwarf and the low-surface brightness late-type galaxies.

The potential of oscillator-type, as it well known, describes one-dimensional particle movement, while the celestial-mechanical potential describes two-dimensional movement in any plane. That is why for the partial case of circular movement along the trajectory $a = const$ it possible decomposes on two independent orthogonal oscillator-type movements. This statement allows find the next relation between mass of particle and parameter $\mu$ in expression (17) - $m^2 = \frac{\mu}{a^3}$, or in the usual units -

$$m = \left(\frac{M}{a^3}\right)^{1/2} \cdot \frac{(G)^{1/2} \cdot \hbar}{c^2}. \qquad (37)$$

For the physical interpretation $\mu$ we assume that it is the ordinary celestial-mechanical parameter. Hence $\mu = GM$, where $G$ is the gravitational constant, $M$ is the central gravitating mass.

Basing on (37) it is possible calculate masses of particles that describe dark matter around dwarf galaxies, for example. Note that for spheroid dwarf $M \approx \frac{4}{3}\pi\tilde{\rho} \cdot a^3$, than $\left(\frac{M}{a^3}\right)^{1/2} \approx 2(\tilde{\rho})^{1/2}$. Here $\tilde{\rho}$ may be interprets as any mean mass density of a dwarf galaxy. As $\frac{(G)^{1/2} \cdot \hbar}{c^2} \approx 8,2 \cdot 10^{-52} \cdot cm^{3/2} \cdot g^{1/2}$, than using the modern typical physical characteristics of spheroid dwarf galaxies $(M \sim (10^7 \div 10^8) M_{Sun}, a \sim 0,3 Kpc)$ [18], we find that $(\tilde{\rho})^{1/2} \sim (10^{-10} \div 10^{-11}) \cdot g^{1/2} \cdot cm^{-3/2}$. Hence, a particle mass is of order $m \sim 10^{-29} eV$. This estimation is in the likelihood correlation with the previous given minimal mass of dark matter's particle $m \approx 10^{-23} eV$.

Moreover, because galaxy's observable characteristics type of sizes is possible measuring with higher accuracy then its gravitating mass, expression (37) allows get possibility for more correct mass estimation. In fact, inversing this calculations we get the following mass for spheroid dwarf galaxies $M \sim (10^{10} \div 10^{11}) M_{Sun}$, i.e. they could be more massive than it is considering now.

**Acknowledgments.** The author thanks professor A.D.Chernin (Sternberg Astronomical Institute, Moscow state University, Moscow, Russia) for his supporting of this idea and his initial acquaintance with the content of this paper.